\documentclass[letterpaper,twocolumn,10pt]{article}
\usepackage{usenix-2020-09}

\usepackage{amsmath}
\usepackage{amssymb}
\usepackage[ruled, linesnumbered, vlined]{algorithm2e}
\usepackage{epsfig}
\usepackage{epstopdf}
\usepackage{etoolbox}
\usepackage{enumitem}
\usepackage{float}
\usepackage{fancyhdr}
\usepackage{graphicx}
\usepackage{graphics}
\usepackage{lscape}
\usepackage{lipsum}
\usepackage{xcolor}
\usepackage{longtable}
\usepackage{tabularx}
\usepackage{multirow}
\usepackage{multicol}
\usepackage{makecell}
\usepackage{listings}
\usepackage[figuresright]{rotating} 
\usepackage{sectsty}
\usepackage{setspace}
\usepackage{subfigure}
\usepackage{breqn}
\usepackage{comment}

\usepackage{url}

\begin{document}

\date{}

\title{\Large \bf FedRings: A Scalable and Topology-Aware Federated Learning Framework for LEO Satellite Constellations}

\author{
{\rm Ziwu Liu, In\^{e}s~Pinto~Gouveia, Rehana Yasmin, Paulo Esteves-Verissimo, Ali~Shoker}\\
CEMSE Division, King Abdullah University of Science and Technology (KAUST)\\
Thuwal 23955-6900, Kingdom of Saudi Arabia
} 



\maketitle

\begin{abstract}

Federated learning over low Earth orbit (LEO) satellite networks is limited by frequent link changes, short contact times, and a highly dynamic topology, making centralized or synchronized training inefficient and hard to scale. To address this, we propose \textit{\textbf{FedRings}}, a decentralized framework that \textit{organizes satellites into ring-based communication structures}. It uses a \textit{spatio-temporal routing strategy} with link-aware communication scheduling to align model exchange with actual visibility windows and time-varying connectivity patterns in LEO. Model updates are propagated along the ring using \textit{adaptive sparse incremental aggregation}, which reduces communication overhead by progressively combining and compressing updates. To handle communication interruptions, a \textit{historical compensation mechanism} maintains training continuity. By combining \textit{topology-aware routing}, \textit{communication scheduling}, and \textit{efficient aggregation}, FedRings enables stable and efficient learning in dynamic LEO networks while reducing communication cost, and experiments show it consistently outperforms existing methods in realistic settings.

\end{abstract}

\section{Introduction}
\label{introduction}

The rapid deployment of low Earth Orbit (LEO) satellite constellations has enabled large-scale space networks composed of hundreds to thousands of satellites. These satellites operate collaboratively and exchange data to support applications such as Earth observation and environmental monitoring. As a result, they continuously generate large volumes of data that must be processed efficiently. In such systems, relying on ground-based processing is inefficient. Large volumes of data must be transmitted to Earth, however, communication is limited by bandwidth and short contact windows. Network links are not always available, and connectivity changes over time. These factors introduce delays and limit centralized processing, motivating moving computation into space and enabling distributed processing within the constellation.

Machine learning is well suited to processing and analyzing satellite data. Federated learning (FL) allows satellites to train models locally and exchange only model updates instead of raw data and to boost privacy. This reduces communication cost and avoids the need to collect all data at a central location. However, applying FL in LEO satellite networks is not straightforward. The network topology is highly dynamic due to the continuous motion of satellites. Connectivity changes over time, and communication links may appear and disappear frequently. Inter-satellite links (ISLs) are often intermittent, and synchronization between nodes is difficult to maintain. These conditions can lead to delayed updates, missing information, and inconsistent model aggregation.

Although connectivity in satellite networks is highly dynamic, the topology itself follows a structured pattern that can be leveraged for more efficient coordination. Satellites are organized into orbital planes, forming ring-like structures. Communication typically occurs along intra-orbit links within the same plane and inter-orbit links between neighboring planes. This organization provides an opportunity to design learning frameworks that account for both spatial and temporal dynamics of the network. However, most existing FL approaches, such as \cite{DSFL, DFedSat, FedLEO, FedSN}, do not explicitly utilize this topology in their design.

This work proposes FedRings, a novel topology-aware FL framework designed for LEO satellite constellations. FedRings aligns model exchange with the multi-ring structure of orbital planes. It uses intra- and inter-orbit links to support structured communication among satellites. It also introduces spatio-temporal routing to select communication paths based on satellite positions and link availability. To improve efficiency, FedRings applies adaptive sparse incremental aggregation to reduce communication overhead. In addition, a historical compensation mechanism is used to mitigate the effect of delayed or missing updates.

By leveraging the inherent topology of the constellation, FedRings enables more efficient and organized model coordination across satellites. It reduces unnecessary communication and avoids dependence on centralized control. FedRings demonstrates how topology-aware design can support scalable and practical FL in dynamic satellite networks. These advancements position FedRings as a critical step toward realizing the full potential of FL in space-based systems. This paper makes the following contributions:

\smallskip
\textit{\textbf{Topology-aware federated learning framework.}} We propose FedRings, a decentralized FL framework designed for LEO satellite constellations. It aligns the learning process with the multi-ring orbital topology instead of assuming generic network connectivity.

\textit{\textbf{Spatio-temporal routing for model exchange.}} A routing approach is introduced that selects communication paths based on satellite positions and link availability over time, enabling efficient and reliable model update propagation.

\textit{\textbf{Adaptive sparse incremental aggregation.}} A communication-efficient method that reduces update size and frequency while maintaining model performance.

\textit{\textbf{Historical compensation mechanism.}} A topology-aware historical compensation mechanism that mitigates delayed and missing updates in multi-ring networks by leveraging spatio-temporal patterns and incremental aggregation.

\textit{\textbf{Rigorous evaluation.}} A detailed evaluation validates that leveraging the multi-ring topology significantly improves communication efficiency, convergence speed, and scalability compared to conventional FL approaches.

\smallskip
The paper is organized as follows. Section~\ref{sec:background} surveys related work. Section~\ref{sec:fedrings} presents the FedRings framework including all components. Section~\ref{sec:eval} describes the evaluation setup, results, comparison with baseline methods. Section~\ref{sec:conclusion} concludes.

\section{Related Work}
\label{sec:background}


Many algorithms have been proposed recently to improve communication efficiency, model consistency, and adaptability in space-based FL systems. This section reviews them, highlighting their key ideas, strengths, and limitations, and how they address LEO challenges such as intermittent connectivity, dynamic topology, and data heterogeneity.

Early works on satellite FL rely on \textit{\textbf{ground-assisted or hybrid architectures}}, which introduce dependency on terrestrial infrastructure. For example, AsyncFLEO (2022) \cite{AsyncFLEO} and FedHAP (2022) \cite{FedHAP} leverage high-altitude platforms (HAPs) to enable asynchronous aggregation and accelerate training, but they still depend on intermediate infrastructure rather than fully exploiting inter-satellite links. Similarly, ground-assisted FL (2022) \cite{FedSat} and FedSpace (2022) \cite{FedSpace} rely on ground stations for coordination and aggregation, \textit{limiting scalability under intermittent connectivity and increasing latency}.

To overcome centralization, several works explore \textit{\textbf{decentralized FL in LEO constellations}}, such as DSFL (2022) \cite{DSFL} and DFedSat (2024) \cite{DFedSat}, which enable direct inter-satellite collaboration. However, these approaches \textit{often assume ideal or simplified communication patterns and do not explicitly optimize topology-aware communication}. FedLEO (2023) \cite{FedLEO} introduces task offloading to enhance resource utilization, but still depends on task redistribution mechanisms that \textit{may incur additional coordination overhead}. Likewise, FedSN (2023) \cite{FedSN} provides a general FL framework for LEO networks, but \textit{lacks detailed handling of dynamic orbital topology and link variability}.

Another line of work focuses on \textit{\textbf{improving communication efficiency and convergence}}. FedGSM (2023) \cite{FedGSM} mitigates gradient staleness in asynchronous FL, while connection-density-aware FL (2024) \cite{ConnectionAwareFL} dynamically adapts aggregation based on satellite–ground connectivity. However, these methods primarily optimize temporal aspects of aggregation and \textit{do not fully exploit structured inter-satellite communication patterns}. FEDMEGA (2024) \cite{Fedmega} further improves communication efficiency by scheduling updates under limited satellite–ground contact windows, yet still \textit{emphasizes satellite–ground interaction rather than intra-constellation cooperation}. In addition, Sparse Incremental Aggregation (2025) \cite{SparseIncrAggSat} introduces incremental in-network aggregation with gradient sparsification, where satellites forward and aggregate updates along intra-orbit links toward a parameter server, improving bandwidth efficiency but still relying on centralized aggregation. General FL optimization methods such as FedUR (2023) \cite{FedUR} improve convergence through adaptive centralized optimizers, but they are not tailored to satellite environments and do not consider dynamic topology or inter-satellite communication constraints.

Several studies address \textit{\textbf{data heterogeneity and multimodal learning}}. FedFusion (2023) \cite{Fedfusion} incorporates manifold learning for multi-satellite and multi-modal data fusion, and ALANINE (2025) \cite{ALANINE} introduces personalized FL to handle non-IID data across heterogeneous satellites. Similarly, OSC-FSKD (2024) \cite{OSC-FSKD} and semi-supervised clustering-based FL (2025) \cite{SemiSupervised-Hi} group satellites based on data similarity to improve convergence. While effective for heterogeneity, these approaches largely \textit{overlook the impact of network topology and routing constraints on learning performance}.

Hierarchical FL frameworks have been proposed to \textit{\textbf{improve scalability and robustness}}. HiSatFL (2025) \cite{HiSatFL} introduces a multi-layer FL architecture with cross-domain privacy adaptation, while RAFL (2025) \cite{RAFL} incorporates reputation-based aggregation to mitigate unreliable participants. Although these methods enhance robustness and trust, they \textit{introduce additional coordination overhead and often assume stable inter-layer communication}.

Recent works also explore \textit{\textbf{security and communication-efficient extensions}}. Blockchain-based FL (2025) \cite{Blockchain} ensures trust and integrity in multi-vendor satellite networks incorporating distributed ledgers, while sat-QFL (2025) \cite{satQFL} integrates quantum key distribution for secure model exchange. In parallel, dataset distillation-based FL (2025)  \cite{Data-Distillation} reduces communication overhead by exchanging compact synthetic data instead of full models. However, these approaches are \textit{largely orthogonal to network structure and do not explicitly optimize communication topology} under dynamic orbital conditions.

Despite significant progress in decentralized FL for LEO satellites, existing approaches largely focus on improving aggregation strategies, communication efficiency, or robustness, while overlooking the role of communication topology. Most frameworks assume simplified or fully connected communication patterns, or rely on ground support, limiting their effectiveness under dynamic orbital conditions and intermittent connectivity. Furthermore, current methods do not fully exploit predictable satellite motion to organize communication and scheduling in a topology-aware manner. These gaps highlight the need for a \textit{\textbf{decentralized, topology-aware approach}} that can leverage orbital structure to improve scalability, communication efficiency, and coordination in FL for LEO constellations, while adapting to irregular connectivity, handling sparse data exchanges, and maintaining robust learning performance under dynamic conditions.

\section{FedRings Framework}
\label{sec:fedrings}

\begin{figure}[h]
    \centering
    \includegraphics[width=\linewidth]{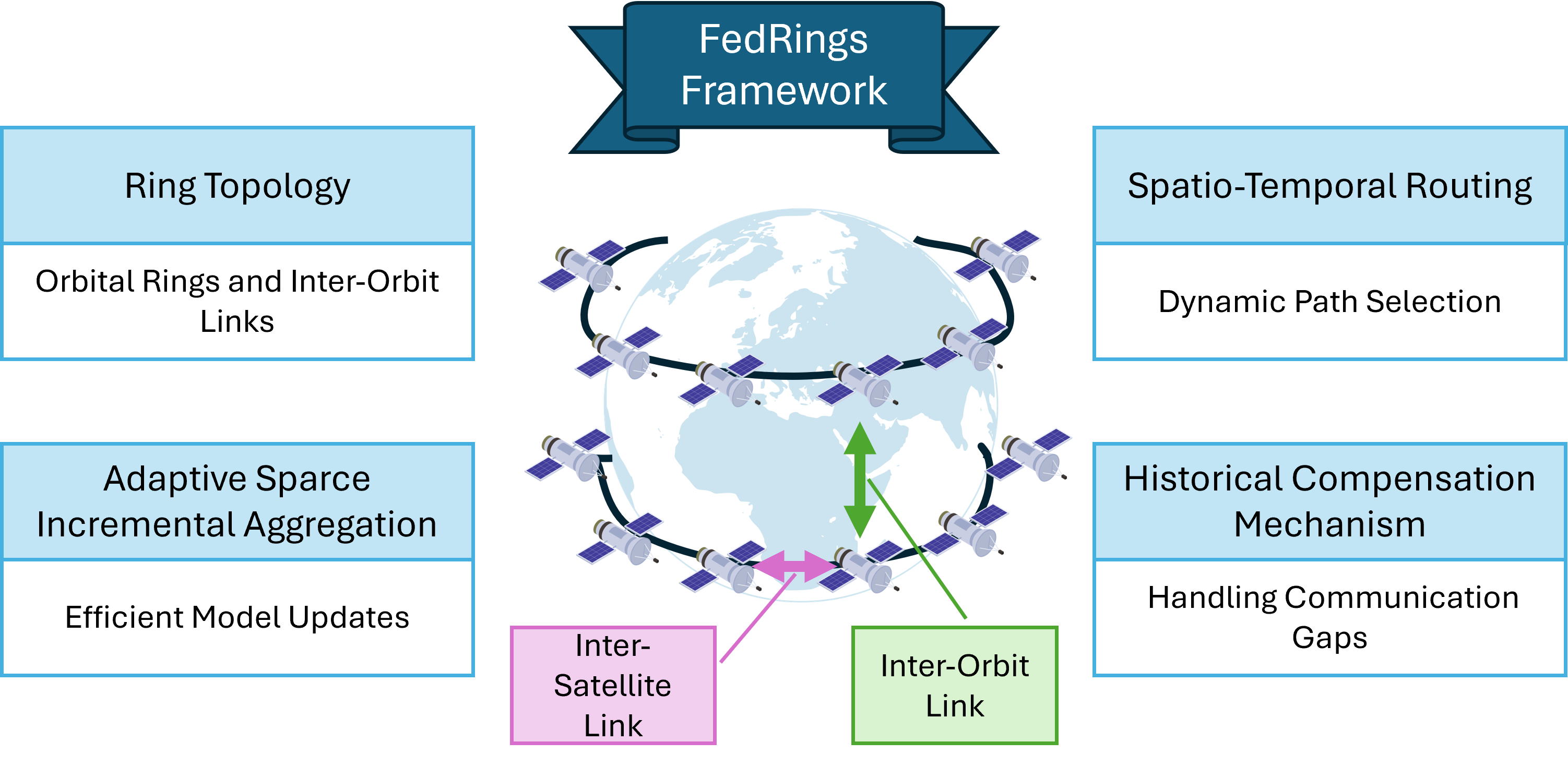}
    \caption{FedRings framework for satellite constellations}
    \label{fig:FedRings}
\end{figure}

To address the aforementioned gaps, we propose \textit{\textbf{FedRings}}, a FL framework designed for LEO satellite constellations. It follows a decentralized approach and uses a \textit{ring-based topology} to structure communication between satellites into rings. A \textit{spatio-temporal routing} mechanism is used to adapt to changing communication windows. Additionally, it includes a \textit{historical compensation mechanism} to handle communication interruptions and ensure robust model convergence. FedRings also incorporates \textit{adaptive sparse incremental aggregation} to reduce communication overhead during model updates, and optimize bandwidth. This design enables efficient communication and supports stable model convergence under dynamic network conditions. Figure~\ref{fig:FedRings} summarizes the main components of the FedRings framework.

\subsection{Ring-Based Topology for FedRings}

\begin{figure}[h]
    \centering
    \includegraphics[width=0.8\linewidth]{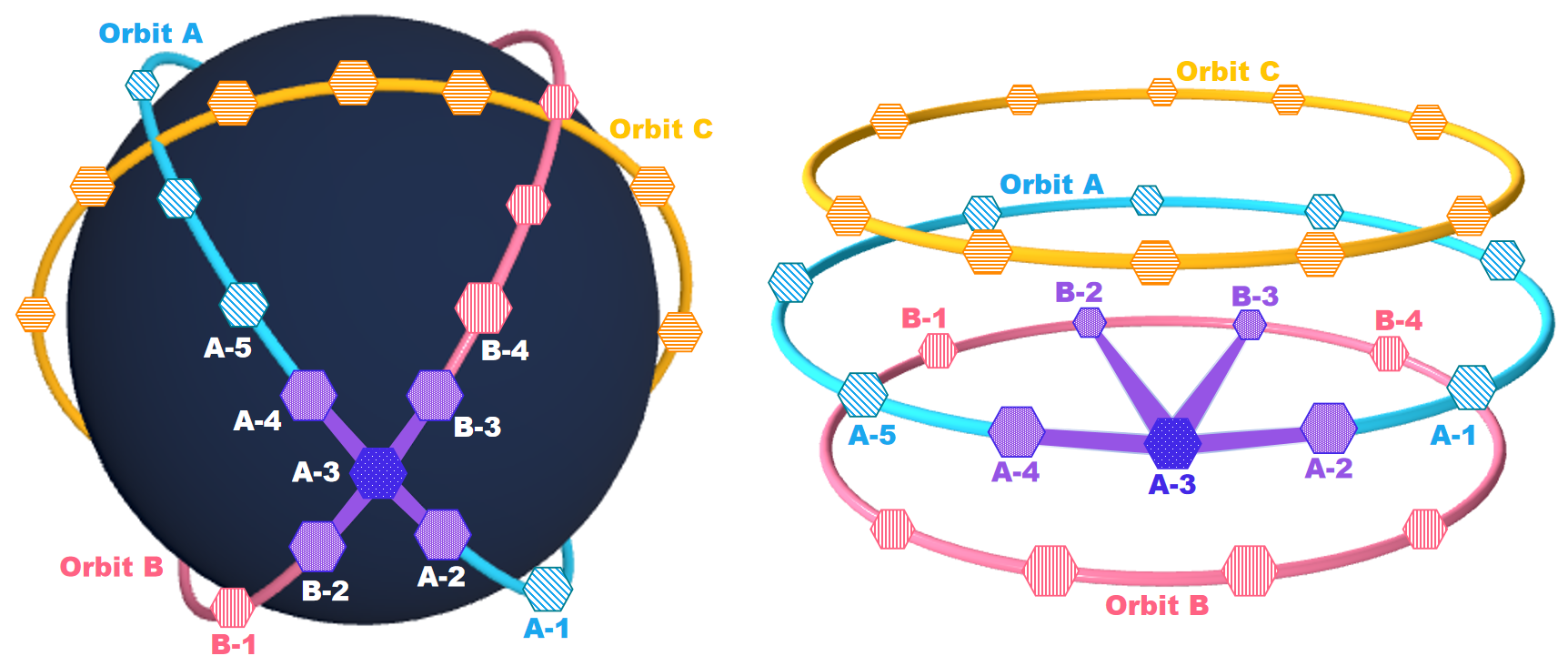}
    \caption{Ring-Based Topology of FedRings.}
    \label{fig:RingTopology}
\end{figure}

The Rings topology is the foundational communication structure employed in the FedRings framework. Inspired by the orbital structure of LEO constellations, FedRings arranges satellites in each orbital plane into a logical ring, where each satellite communicates with its two immediate neighbors via inter-satellite links (ISL) -- one upstream and one downstream. These bidirectional links facilitate seamless data exchange for FL tasks, particularly during the all-reduce operation. In addition, satellites in different rings can communicate during inter-orbit communication windows, allowing data to flow across orbital planes, thus forming a dynamically connected ``Rings'' topology, as shown in Figure~\ref{fig:RingTopology}

In the FedRings framework, the ring-based topology enables efficient and decentralized execution of distributed learning by supporting structured data aggregation, reducing communication overhead while maintaining robustness and scalability. This is achieved through coordinated intra- and inter-ring communication, which facilitates parameter exchange for distributed federated learning.

\subsection{Spatio-Temporal Routing}

The \textit{spatio-temporal routing} strategy in FedRings is a dynamic path selection method that enables efficient communication and coordination under the intermittent connectivity of LEO satellite constellations. It addresses the unique dynamics of these constellations, where satellites move at high velocities relative to one another, causing communication windows to open and close rapidly. As a result, the relative positions of satellites continuously change, leading to a dynamically evolving multi-ring topology driven by irregular communication opportunities. Therefore, FedRings does not rely on exact spatial relationships between satellites and instead focuses on whether a communication window exists between them, allowing each satellite to dynamically select the optimal communication path within each available window.

Implementing a spatio-temporal routing strategy presents several challenges. The fast movement of LEO satellites creates short and intermittent communication windows, so accurate scheduling is needed to use each opportunity. Frequent position changes make paths unstable, as links form and break quickly, requiring adaptive selection to maintain connectivity. Limited bandwidth and power further restrict data transmission, so paths must maximize throughput within each short window. To address these challenges, FedRings uses predictive modeling and adaptive scheduling. Predictive models, through simulation, estimate communication windows in advance and provide each satellite with a communication schedule. Adaptive scheduling then prioritizes these windows based on link duration, data load, and resource limits, helping each satellite make better use of its connection time.

\begin{figure}
    \centering
    \includegraphics[width=\linewidth]{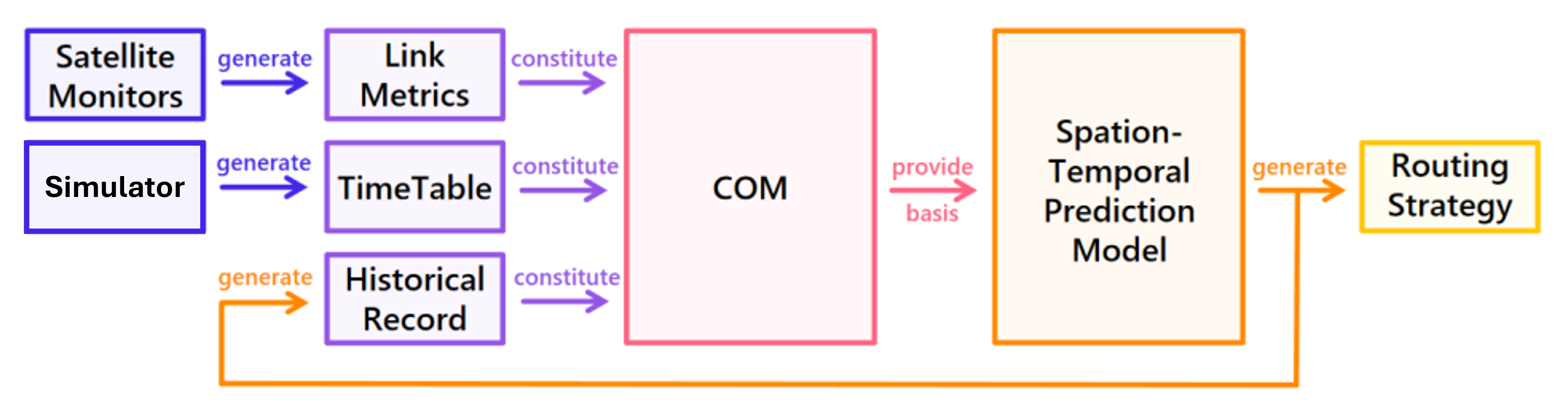}
    \caption{Spatio-temporal routing strategy}
    \label{fig:routing_flowchart}
\end{figure}

As illustrated in Figure~\ref{fig:routing_flowchart}, the \textit{\textbf{Routing Strategy}} follows a structured process based on three main inputs: \textit{TimeTable}, \textit{Link Metrics}, and \textit{Historical Records}. A satellite simulator such as Systems Tool Kit (STK) generates a pre-simulated \textit{\textbf{TimeTable}}, a topology-aware dataset of communication windows between satellites. It considers orbital motion, velocity, and communication range to determine when inter-satellite links are available and which nodes can exchange model updates in a multi-ring setup. It defines active intra- and inter-ring links and the duration of each communication opportunity for the FL process. The \textit{\textbf{Link Metrics}} are generated by the communication monitors which record the state of links between each satellite representing the quality of the communication. The \textit{\textbf{Historical Records}} store system states, such as failures or overload. Together, these inputs form the \textit{\textbf{Communication Opportunity Matrix}}, which feeds into a \textit{\textbf{Spatio-Temporal Prediction Model}} that analyzes communication patterns and selects suitable routing strategies, making it central to the adaptive communication strategy. By combining pre-simulated data with real-time feedback, each satellite can adapt its communication decisions and use resources efficiently in a dynamic LEO environment. The \textit{Routing Strategy} is continuously updated using feedback stored in the \textit{Historical Record}, creating a loop that improves performance over time. This process supports reliable and efficient communication across the constellation.

\subsubsection{Communication Opportunity Matrix}


Communication Opportunity Matrix (COM) serves as the foundation for dynamic communication decision-making. In COM, the full constellation is modeled as a set of interconnected rings \( G(t) = \bigcup_{k=1}^{M} G_k(t) \), with inter-plane communication defined by additional inter-ring links \( E_{inter}(t) \) between satellites in different planes. This connectivity is captured by a time-dependent adjacency matrix \( A(t) \), where each entry indicates the presence of an active communication link based on the TimeTable:

\begin{equation}\label{eq:adjacency_matrix}
    A_{ij}(t) = \begin{cases} 
    1, & \text{if satellites } i \text{ and } j \text{ can communicate at time } t \\
    0, & \text{otherwise}
    \end{cases}
\end{equation}

This matrix changes over time according to the available communication windows, allowing the network structure to dynamically adapt to real-time connectivity. For any two satellites $i$ and $j$ within the same orbital plane $G_k$ , the intra-orbital data rate $C_{intra} (i, j, t)$ is valid only during the TimeTable-defined communication window:

\begin{equation}\label{eq:intra_data_rate}
    C_{intra}(i, j, t) = B_{intra} \log_2 \left(1 + \frac{P_{rx,ij}^{intra}}{N_0}\right)
\end{equation}

where \( B_{intra} \) is the intra-orbital channel bandwidth, \( P_{rx,ij}^{intra} \) is the received power, and \( N_0 \) represents the noise spectral density. This formula is applicable only during times when the TimeTable indicates an active communication window. For satellites in different planes, the inter-orbital data rate \( C_{inter}(i, j, t) \) similarly depends on the TimeTable:

\begin{equation}\label{eq:inter_data_rate}
    C_{inter}(i, j, t) = B_{inter} \log_2 \left(1 + \frac{P_{rx,ij}^{inter}}{N_0}\right)
\end{equation}

where \( B_{inter} \) is the inter-orbital bandwidth and \( P_{rx,ij}^{inter} \) represents the inter-plane received power. Here again, the communication rate is defined only for intervals where the TimeTable designates a communication opportunity. The TimeTable's communication windows impact both latency and path selection. The total latency $\tau_{total}$ for data transfer between satellites $i$ and $j$ is calculated by considering only the hops available within the active windows:

\begin{equation}\label{eq:latency}
    \tau_{total} = \sum_{p=1}^{P} \left( \frac{d_{p}}{c} + \tau_{proc} \right)
\end{equation}

where $P$ represents the number of hops in the path, $d_p$ is the distance for each hop defined by the adjacency matrix $A(t)$ during active windows, $c$ is the speed of light, and $\tau_{proc}$ is the processing delay. This dynamic adjustment ensures the latency calculation reflects only feasible paths, adhering to the TimeTable's constraints. Routing decisions are based on a time-dependent routing matrix $R(t)$ derived directly from the \begin{equation}\label{eq:routing_matrix}
    R_{ij}(t) = \begin{cases}
    1, & \text{if satellite } i \text{ can relay data to } j \text{ during } t \\
    0, & \text{otherwise}
    \end{cases}
\end{equation}

This matrix enables adaptive routing, ensuring each path is feasible only within the designated communication windows.

Network resilience is reinforced by the connectivity K, defined by the TimeTable's availability of disjoint paths:
\begin{equation}\label{eq:connectivity}
    K = \min_{s,t} \# \text{ of disjoint paths from } s \text{ to } t
\end{equation}
Maximizing $K$ through multiple communication windows enhances network robustness, as disjoint paths provide redundancy in case of temporary link loss. 

Each satellite uses TimeTable data to initialize its COM, allowing it to predict future communication opportunities and window durations. As a static representation of potential communication paths, the TimeTable provides a foundation for pre-computing routes and transmission plans, enhancing the system's initial communication efficiency. However, COM is not merely a static schedule. It updates based on real-time link conditions. Each satellite monitors signal strength, delay, and link quality during active communication windows. When a link performs poorly, the satellite lowers its priority in the COM or marks it as "unavailable". This helps avoid unreliable paths and keeps transmission stable. If a link fails, the satellite switches to a backup path using multi-path routing.

In addition to real-time adjustments, COM management in FedRings incorporates historical optimization. After each transmission, satellites update path priorities based on link performance. Reliable paths are preferred, while weak ones are reduced or removed. Over time,  as historical data accumulates, COM transitions from a static matrix derived from the TimeTable to a dynamic, adaptive structure by combining real-time feedback with historical data. This improves routing and resource use in dynamic networks. All decisions are made locally by each satellite. They use the TimeTable as a guide but adapt to current conditions. No central control is needed. This avoids single points of failure and supports scalability in large, changing constellations.

\subsubsection{Spatio-Temporal Prediction Model}


The spatio-temporal prediction model works with a dynamic multi-path routing strategy, allowing each satellite to predict and manage its communication opportunities independently. To support this, a Star Walker-type constellation is modeled within the prediction framework as a structured multi-ring network, where each orbital plane forms a separate ring. This representation captures the regular orbital pattern while enabling effective routing decisions across and within rings. This constellation is defined by $M$ orbital planes, each containing $N_{k}$ satellites in the $k$-th plane ($k \in {1, 2, . . . , M }$). Communication between satellites is organized based on the communication windows outlined in the TimeTable. The intra-orbital communication within each plane $k$ is represented by a dynamic graph $G_{k} (t) = (V_{k} , E_{k} (t))$, where $V_{k}$ denotes satellites in the orbital plane and $E_{k} (t)$ represents the active links within the plane at time $t$, as indicated by the TimeTable. Communication links are only established when a communication window is open, ensuring accurate simulation of real-world connectivity. During route analysis, the TimeTable is combined with this graph to guide decisions. The following sections outline the core mechanisms and their integration.



\medskip
\noindent \textbf{Multi-Path Selection and Rapid Failure Response:} The dynamic multi-path routing mechanism leverages Yen's Algorithm (see Algo~\ref{algo:algo1}) to pre-compute and rank multiple feasible paths for communication using the TimeTable, a matrix $G = (V, E)$ that encodes detailed information about all feasible communication opportunities between satellites. The algorithm identifies the top $k$ shortest paths between a source satellite $s$ and a target satellite $t$, ensuring that the selected paths optimize for COM. By iteratively computing alternative paths while avoiding overlapping or redundant segments, this approach enables satellites to dynamically select from precomputed paths to maintain continuous data flow and mitigate disruptions caused by link failures or quality degradation.

The routing process starts by selecting the optimal path $P_1$ from precomputed paths. This path is obtained using Dijkstra's algorithm on graph $G$, based on TimeTable edge weights. If the primary path fails or degrades, the system selects the next-best path $P_i$ from the set $K_{\text{paths}}$ of ranked $k$ shortest paths. Alternative paths are built by combining a root segment $\text{Root} = (v_1, v_2, \dots, v_j)$ with a spur path $\text{Spur}$ from node $v_j$ to $t$. This provides diverse and efficient backup paths. When a link becomes unresponsive, the satellite quickly switches to another precomputed path, reducing disruption and keeping communication stable in dynamic LEO networks. By integrating Yen's Algorithm, FedRings improves resilience and enables real-time routing adaptation. It also increases reliability and reduces reliance on centralized control.

\begin{algorithm}[!htbp]
\caption{Yen's Algorithm-Based Multi-Path Selection with Explanations}
\label{algo:algo1}
\KwIn{$G = (V, E)$: Communication graph with vertices $V$ and edges $E$;\\
$s$: Source satellite; $t$: Target satellite;\\
$k$: Number of desired paths;\\
$\text{TimeTable}$: Precomputed communication opportunities (edge weights);\\
$\text{QoS\_metric}$: Quality metric (e.g., signal strength, delay).}
\KwOut{$K_{\text{paths}}$: Top $k$ shortest paths from $s$ to $t$.}

\textbf{Step 1: Initialization}\;

\textbf{initialize} $K_{\text{paths}} \gets \emptyset$\; 
$\text{candidate\_paths} \gets \emptyset$\; 

\textbf{Step 2: Find the first shortest path}\;

$P_1 \gets \text{Dijkstra}(G, s, t, \text{TimeTable})$ \\ 
\textit{// Find the shortest path using Dijkstra's algorithm based on edge weights in $\text{TimeTable}$}\;

Add $P_1$ to $K_{\text{paths}}$ \\ 
\textit{// Store the first shortest path in the result set}\;

\textbf{Step 3: Iteratively find remaining $k-1$ paths}\;
\For{$i \gets 2$ \KwTo $k$}{ 
    \For{$j \gets 1$ \KwTo $\text{len}(P_{i-1}) - 1$}{ 
        $\text{Root} \gets P_{i-1}[1:j]$ \\ 
        \textit{// Take the first $j$ nodes of $P_{i-1}$ as the root path $\text{Root} = (v_1, v_2, \dots, v_j)$}\;

        $\text{Temp\_G} \gets G$ \\ 
        \textit{// Create a temporary graph by removing overlapping edges}\;
        
        \ForEach{$P \in K_{\text{paths}}$}{ 
            \If{$P$ starts with $\text{Root}$}{
                Remove $P[j]$ from $\text{Temp\_G}$ \\ 
                \textit{// Ensure alternative paths do not reuse this segment.}
            }
        }

        $\text{Spur} \gets \text{Dijkstra}(\text{Temp\_G}, \text{Root}[j], t, \text{TimeTable})$ \\ 
        \textit{// Find the shortest path from spur node $v_j$ to $t$ in the modified graph.}

        $\text{NewPath} \gets \text{Root} + \text{Spur}$ \\ 
        \textit{// Merge $\text{Root}$ and $\text{Spur}$ to form a candidate path.}\;
        
        Add $\text{NewPath}$ to $\text{candidate\_paths}$\;
    }

    Sort $\text{candidate\_paths}$ by $\text{QoS\_metric}$ \\ 
    \textit{// Rank all candidate paths based on the quality metric.}\;
    
    $\text{BestPath} \gets \text{candidate\_paths}[1]$ \\ 
    \textit{// Choose the top-ranked path.}\;
    
    Add $\text{BestPath}$ to $K_{\text{paths}}$\; 
    
    Remove $\text{BestPath}$ from $\text{candidate\_paths}$\;
}

\textbf{Step 4: Return Result}\;

\textbf{return} $K_{\text{paths}}$ \\ 
\textit{// Output the top $k$ shortest paths for dynamic routing.}

\end{algorithm}

\medskip 
\noindent \textbf{Real-Time Communication Window Monitoring and Scheduling:} 
Complementing multi-path routing, FedRings uses decentralized communication window monitoring and scheduling. Each satellite uses the TimeTable to predict upcoming communication windows, removing the need for centralized coordination and enabling autonomous operation in dynamic LEO environments. During active windows, it monitors link quality, such as signal strength, delay, and packet loss, and adapts transmission accordingly. For example, if a link degrades but is still usable, the satellite reduces the data rate and prioritizes critical updates to make better use of limited time. If the link fails, it switches to a backup path, ensuring continuous data transfer. As communication windows approach, satellites prepare data packets and adjust routing based on the predicted time and bandwidth in the TimeTable. This proactive scheduling reduces missed opportunities and improves resource use. By combining prediction with real-time adjustment, the system balances planning and adaptation.

\begin{figure}[htp]
    \centering
    \includegraphics[width=0.33\textwidth]{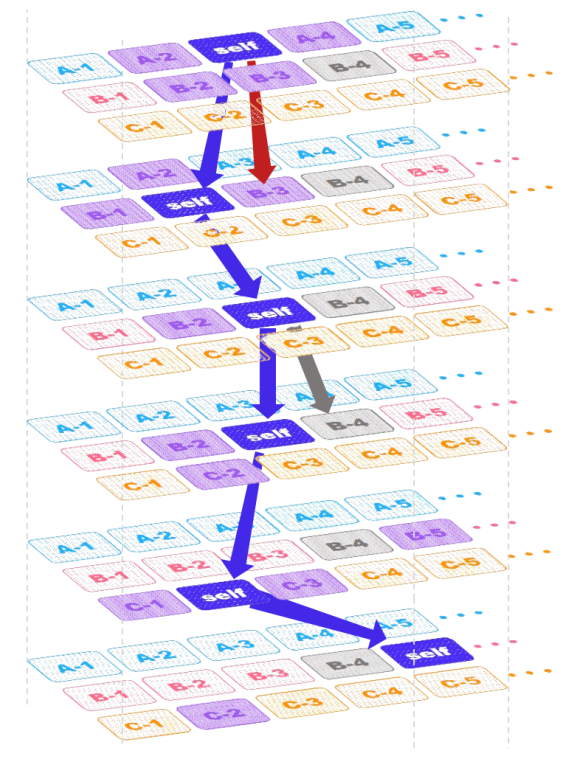}
    \caption{Spatio-Temporal routing strategy base on COM}
    \label{fig: rount}
\end{figure}

Let $A$, $B$, and $C$ be orbits, as shown in the Figure~\ref{fig: rount}, with time progressing from top to bottom. The goal is to send a data packet from $A-3$ to $B-5$. The blue arrows show the selected route. The packet makes three inter-orbit hops, one intra-orbit hop, and one waiting step. At first, sending from $A-3$ to $B-3$ and then along orbit $B$ to $B-5$ seems optimal during the first COM. However, this link is avoided due to poor quality (as indicated by the red arrow). At $B-3$, forwarding via $B-4$ is also skipped, as $B-4$ is marked offline in historical records (as indicated by the grey arrow). Instead, the packet is routed through $C-2$ as an intermediate relay and then delivered to $B-5$. In space, such unpredictable situations are common. The spatio-temporal routing algorithm handles them by adapting decisions using real-time link conditions and historical records. This ensures successful packet delivery even in adverse conditions.

Multi-path selection and real-time communication window monitoring form the core of the FedRings routing strategy. Multi-path routing provides redundancy and handles link failures and quality changes by switching paths when needed. Real-time monitoring adapts transmission based on current link conditions. The TimeTable adds predictive scheduling, improving resource use during short communication windows. By relying on precomputed paths, the system also reduces computation while maintaining high performance.

\subsection{Adaptive Sparse Incremental Aggregation}

\begin{figure}
    \centering
    \includegraphics[width=\linewidth]{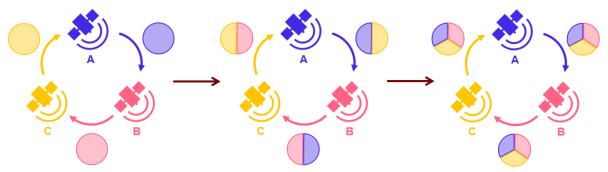}
    \caption{Incremental aggregation example with 3 satellites.}
    \label{fig:incremental_aggregation}
\end{figure}

Adaptive Sparse Incremental Aggregation (ASIA) is introduced for FL over ring-based satellite networks to reduce communication overhead while preserving model quality. Instead of exchanging full model updates with multiple peers, ASIA uses incremental aggregation, where updates are progressively combined as they move along the ring. Each satellite receives an aggregated update from its predecessor, merges it with its local update, and forwards the result to the next node. A key design element of ASIA is sparsification applied in this process, keeping only the most significant components and using a fixed-size message structure. This reduces communication cost while preserving the most significant components of the model updates at each step. In this way, ASIA extends incremental aggregation for spatio-temporal routing in LEO ring topologies, enabling scalable and bandwidth-efficient collaborative learning across satellites. For example, in Figure~\ref{fig:incremental_aggregation}, a ring consists of three satellites $\mathcal{A}$, $\mathcal{B}$, and $\mathcal{C}$. $\mathcal{A}$ sends its update to $\mathcal{B}$, which merges it with its own update and forwards the result to $\mathcal{C}$. Thus, $\mathcal{C}$ receives an already aggregated update without direct communication from $\mathcal{A}$. ASIA also uses Top-$Q$ sparsification to keep only the most important parameters, ensuring constant transmission size across rounds.

The ASIA process works as follows. Each satellite receives the aggregated update, merges it with its local update, and applies Top-$Q$ sparsification to retain a fixed number of non-zero gradients. The result is then forwarded to the next hop. A time-correlated sparsification strategy with a global mask is also used to keep stable parameters across rounds and reduce unnecessary changes. This improves communication efficiency and supports faster convergence. Finally, when the update reaches the central node, weighted aggregation of all contributions completes the model update for that round.

\subsubsection{ASIA Design}

In ASIA, a time-correlated sparsification strategy is used to improve communication efficiency and support stable model convergence across aggregation rounds. It leverages historical sparsification patterns so that each satellite consistently transmits important parameters. This is well suited to the FedRings ring topology, where satellites communicate sequentially and pass updates around the ring. Keeping key parameters consistent reduces redundant transmissions and improves efficiency.

The core idea is a global sparsification mask. This mask selects a fixed set of important parameters that remain stable over time. In FedRings, these parameters can be identified from repeated gradient significance across satellites or their impact on convergence. During each transmission, the mask is shared across all nodes. When a satellite receives an update, it applies the global mask, merges it with its local update, and forwards the result to the next node. This maintains consistency in parameter selection across the ring. To add flexibility, ASIA also allows local adaptation. Each satellite can include a small number of additional high-priority parameters based on its local gradients. For example, if a satellite detects sudden local changes, it can add those updates to the transmission. These local updates pass through the ring without affecting the stability provided by the global mask. This balances global consistency with local responsiveness in FedRings.

This design improves both stability and efficiency. The same key parameters are preserved across rounds, which helps model convergence and reduces fluctuations in communication. It also removes unnecessary parameter changes, improving bandwidth usage under limited communication budgets. The combination of a stable global mask and adaptive local updates makes FedRings more efficient in dynamic and bandwidth-constrained satellite networks. To implement this, the global mask must be synchronized across all nodes in FedRings. A coordinating node updates and distributes the mask every few rounds. Each satellite uses the same mask during sparsification, ensuring consistency across the ring. Efficient protocols also handle local mask adjustments to avoid network overload. Regular updates allow the mask to adapt to changes in model behavior and training data while keeping communication aligned.

The time-correlated sparsification enables FedRings to transmit key parameters consistently while still adapting to local conditions. This reduces communication cost and improves convergence stability. This design provides clear advantages in real-world applications, especially in bandwidth-limited satellite constellations, making FL more efficient and stable in complex distributed environments.

\subsubsection{Scalability of ASIA}

ASIA is designed for large-scale distributed networks, such as satellite constellations with tens or hundreds of nodes. It remains scalable through three key design features.

\medskip 
\noindent \textbf{Constant-Length Design:} ASIA uses a constant-length sparsification strategy, where each transmission carries a fixed number of non-zero gradient parameters, regardless of network size. This keeps communication cost stable and bandwidth usage predictable. In large constellations, traditional aggregation increases data load as more nodes join. In contrast, ASIA avoids this growth by keeping transmission size fixed. This is important in FedRings, where bandwidth and link quality are limited. It ensures stable communication even at large scale deployments.

\medskip 
\noindent \textbf{Incremental Aggregation and Ring Adaptation:} ASIA uses incremental aggregation in the FedRings ring topology. Each satellite communicates only with its immediate neighbors, receives an aggregated update, merges it with its local update, and forwards it to the next node. This keeps communication local and avoids network-wide exchanges. As the network grows, each node still interacts with only two neighbors. For example, in a FedRings network with 100 satellites, each node communicates only with its adjacent nodes. This prevents bottlenecks and supports scalable deployment.

\medskip 
\noindent \textbf{Simplified Synchronization Strategy:} To reduce synchronization overhead, the global sparsification mask is updated periodically instead of every round. Nodes use the same mask between updates, which reduces coordination cost. Additionally, a coordinating node or ground station can manage and distribute the mask across the network. This periodic update strategy maintains consistency while avoiding frequent synchronization delays, making ASIA suitable for large-scale FedRings deployments.

\subsection{Historical Compensation Mechanism}

Communication in FedRings is often affected by space environment conditions, which can lead to missing updates. To handle this, a historical compensation mechanism is introduced. It uses stored historical parameters at each satellite, together with local updates, to reconstruct missing information and maintain continuous aggregation. The goal is to reduce the impact of communication loss on training. When a satellite does not receive expected updates from its neighbors in the ring, it reconstructs the missing values using stored history. This avoids retransmission and keeps the aggregation process running. It also allows satellites to continue participating even during link interruptions, improving system robustness.

\medskip
\noindent \textbf{Historical Parameter-Based Compensation.} 
The historical parameter-based compensation mechanism relies on neighbor quality in FedRings. Satellites with stable links are treated as high-quality neighbors, while others are low-quality. Stability is determined using the COM. A neighbor $j \in N$ is classified as high-quality if COM indicates a stable link COM.isStableLink($i, j$), with at least 80\% success over the last 5 rounds (see Algorithm~\ref{algo:algo2}). These neighbors form the set $N_{high}$ and are used for reliable historical storage, as their stable connectivity ensures the reliability of stored parameters $w_j$. Low-quality neighbors form the set $N_{low}$, often due to intermittent links or cross-orbit communication limits. The classification is updated dynamically. A high-quality neighbor is downgraded after three consecutive failures, while a low-quality neighbor is upgraded when stability improves.

\setcounter{AlgoLine}{0}

\begin{algorithm}[ht]
\caption{Classify Neighbors Based on COM}
\label{algo:algo2}
\KwIn{Neighbor set $\mathcal{N}$, Communication Opportunity Matrix (COM)}
\KwOut{High-quality neighbor set $\mathcal{N}_{\text{high}}$, Low-quality neighbor set $\mathcal{N}_{\text{low}}$}

\textbf{initialize} $\mathcal{N}_{\text{high}} \gets \emptyset$\; 
$\mathcal{N}_{\text{low}} \gets \emptyset$\;

$\mathcal{N}_{\text{high}} \gets \emptyset$\; 
$\mathcal{N}_{\text{low}} \gets \emptyset$\; 

\ForEach{$j \in \mathcal{N}$}{ 
    \If{$\text{COM.isStableLink}(i, j)$}{
        Add $j$ to $\mathcal{N}_{\text{high}}$\; 
        Remove $j$ from $\mathcal{N}_{\text{low}}$\; 
    }
    \Else{
        Add $j$ to $\mathcal{N}_{\text{low}}$\; 
        Remove $j$ from $\mathcal{N}_{\text{high}}$\; 
    }
}
\textbf{return} $\mathcal{N}_{\text{high}}, \mathcal{N}_{\text{low}}$\;

\end{algorithm}

During \textit{incremental aggregation}, each satellite $i$ combines updates $u_j \in U$ from its neighbors $j \in N$. If an update is received, it is added to the aggregation $w_{agg}$, and for high-quality neighbors, it is also stored as historical parameter $w_j$:

\[
\mathbf{w}_{\text{agg}} \gets \mathbf{w}_{\text{agg}} + \mathbf{u}_j, \quad \text{if } j \in \mathrm{N}_{\text{high}}, \mathbf{w}_j \gets \mathbf{u}_j
\]

If an update is missing ($u_j = \emptyset$), compensation is applied (see Algorithm~\ref{algo:algo3}). For $j \in N_{high}$, the stored historical value $w_j$ is used as compensation ($w_{comp} \leftarrow w_j$). For $j \in N_{low}$, the satellite uses its current aggregated state $w_{agg}$ as a fallback ($w_{comp} \leftarrow w_{agg}$). This dual strategy ensures stable aggregation even under link failures, while avoiding over-reliance on unreliable neighbors, maintaining the balance between accuracy and efficiency in FL.

\setcounter{AlgoLine}{0}

\begin{algorithm}[!ht]
\caption{Historical Compensation Process}
\label{algo:algo3}
\KwIn{Communication updates $\mathcal{U}$, Neighbor sets $\mathcal{N}_{\text{high}}$, $\mathcal{N}_{\text{low}}$, Historical parameters $\mathbf{w}_j$}
\KwOut{Aggregated model parameters $\mathbf{w}_{\text{agg}}$}

\textbf{initialize} $\mathbf{w}_{\text{agg}} \gets \mathbf{w}_0$ \; 

\ForEach{$j \in \mathcal{N}$ with update $\mathbf{u}_j \in \mathcal{U}$}{
    \If{$\mathbf{u}_j \neq \emptyset$}{ 
        \If{$j \in \mathcal{N}_{\text{high}}$}{
            $\mathbf{w}_j \gets \mathbf{u}_j$\; 
        }
        $\mathbf{w}_{\text{agg}} \gets \mathbf{w}_{\text{agg}} + \mathbf{u}_j$\; 
    }
    \Else{ 
        \If{$j \in \mathcal{N}_{\text{high}}$}{
            $\mathbf{w}_{\text{comp}} \gets \mathbf{w}_j$ \; 
        }
        \Else{
            $\mathbf{w}_{\text{comp}} \gets \mathbf{w}_{\text{agg}}$ \; 
        }
        $\mathbf{w}_{\text{agg}} \gets \mathbf{w}_{\text{agg}} + \mathbf{w}_{\text{comp}}$\; 
    }
}

\textbf{return} $\mathbf{w}_{\text{agg}}$\;

\end{algorithm}

\medskip
\noindent \textbf{Impact on Convergence and Stability.}
The historical compensation mechanism improves model convergence in FedRings by handling missing updates during aggregation. When communication fails, missing parameters can introduce bias in the model  and slow or destabilize training. By using historical parameters to reconstruct missing values, the mechanism preserves model integrity. This allows each node to contribute reliable updates even under occasional failures, leading to smoother and more consistent convergence.

It also improves system stability in FL. By generating substitute values for missing data, nodes can continue aggregation without interruption. This prevents stalled updates in the ring topology, which is important in satellite networks with unstable links. As a result, the system remains robust and continues operating even under unreliable conditions.
\section{Evaluation}
\label{sec:eval}

As the constellation type with the strongest structural compatibility, the Walker Star constellation is the most widely applied. Therefore, this paper chooses to base the simulation experiments on the Walker Star constellation, which we model in the STK simulator. Moreover, since the TimeTable is generated from simulation results, changes in constellation type do not affect the structure of FedRings but only alter the outcomes of the Spatio-Temporal Routing. In the scenario constructed in this paper, each orbital plane
contains 15 satellites, which are evenly and equidistantly distributed. The entire constellation consists of six such orbital planes, which share the same inclination and are evenly spaced. This is denoted as $Walker(t = 6/p = 90/f = 1)$ or $Walker(6/90/1)$. Here, $t$ represents the total number of orbital planes. $p = n \times t$ represents the total number of satellites ($n$ is the number of satellites per orbital plane). $f$ is the phasing factor, indicating the phase difference between adjacent orbital planes, where $f = 1$ signifies evenly spaced orbital planes.

The essential and indispensable key parameters for constructing the STK satellite simulation environment are as follows: $N_C$: The total number of satellite constellations in the system. Each constellation is a group of satellites in the same orbit; $N_S$: The number of satellites within each constellation. These satellites are connected in a ring topology within the constellation; $D_{max}$: The maximum distance at which satellites from different constellations can communicate with each other when they are close enough; $N_G$: The number of ground servers in the system; $G_{latitude}$: The location latitude where the ground server is situated, which can handle higher performance operations such as global model aggregation; $\theta_G$: The angular range within which the ground server can observe and communicate with the satellites. This angle determines the visibility of the satellites from the ground station; $I_C$: The inclination of satellites in the constellation. It determines the geographic area that this constellation can cover; $H_{LEO}$: Refers to the typical altitude range for satellites in LEO, which is generally between 500 km and 2000 km above the Earth's surface. The value ranges and default values of these parameters used in this project are presented in Table~\ref{tab:param_settings}.

\begin{table}[htp]
    \centering
    \renewcommand{\arraystretch}{1.0} 
    \setlength{\tabcolsep}{4pt} 
    \small
    \begin{tabular}{p{2.2cm}p{2.5cm}p{2.2cm}} 
        \Xhline{1.2pt}
        \multicolumn{1}{c}{\textbf{Parameters}} & \multicolumn{1}{c}{\textbf{Range}} & \multicolumn{1}{c}{\textbf{Default}} \\ 
        \Xhline{1.2pt}
        \makecell*[c]{$N_C$} & \makecell*[c]{3$\sim$9} & \makecell*[c]{5} \\
        \hline
        \makecell*[c]{$N_S$} & \makecell*[c]{11$\sim$20} & \makecell*[c]{12} \\
        \hline
        \makecell*[c]{$D_{max}$} & \makecell*[c]{$5000\sim7500km$} & \makecell*[c]{5000km} \\
        \hline
        \makecell*[c]{$N_G$} & \makecell*[c]{1$\sim$3} & \makecell*[c]{1} \\
        \hline
        \makecell*[c]{$G_{latitude}$} & \makecell*[c]{$-90^\circ \sim 90^\circ$} & \makecell*[c]{$21.3^\circ$} \\
        \hline
        \makecell*[c]{$\theta_G$} & \makecell*[c]{$10^\circ \sim 90^\circ$} & \makecell*[c]{$40^\circ$} \\
        \hline
        \makecell*[c]{$I_C$} & \makecell*[c]{$30^\circ \sim 65^\circ$} & \makecell*[c]{$65^\circ$} \\
        \hline
        \makecell*[c]{$H_{LEO}$} & \makecell*[c]{$340\sim1200km$} & \makecell*[c]{$550km$} \\
        \hline
        \Xhline{1.2pt}
    \end{tabular}
    \caption{Parameters Setting}
    \label{tab:param_settings}
\end{table}

In order to verify the performance of FedRings, we select models and datasets that fit the real requirements of the constellation as much as possible. Such model size and dataset characteristics are consistent with the characteristics of application scenarios, so it has a reliable reference.

\medskip
\noindent \textbf{Model:} FedRings selects DenseNet-121 as the model. DenseNet-121 is a convolutional neural networks (CNNs) architecture designed for image classification and other tasks like segmentation. In decentralized federated learning, nodes may fail to fully synchronize models due to link interruptions or data loss. DenseNet-121 is more robust in such scenarios because of its feature reuse and shorter gradient paths, which reduce the gradient vanishing problem. Additionally, DenseNet-121 has around 8 Mb parameters, resulting in lower communication and storage overhead, making it suitable for communication-constrained federated learning scenarios. The learning rate is set to 0.1, the optimizer used is SGD, and the weight decay is set to 0.001. The training epochs are set to 300 or 350, depending on the dataset.

\medskip
\noindent \textbf{Dataset:} To highlight the performance of FedRings in satellite constellation environments, this study specifically uses three datasets derived from satellites: EuroSAT~\cite{EuroSAT}, So2Sat~\cite{So2Sat}, and DeepGlobe~\cite{DeepGlobe}. \textit{\textbf{EuroSAT}} is a dataset containing 10 classes of Earth observation images captured by Sentinel-2 satellites. The image size is 64×64, with an actual input size of 64×64; \textit{\textbf{So2Sat}} is a multispectral remote sensing dataset with 42 classes of Earth cover types. The original image size is 32×32, interpolated to 224×224 for input; and \textit{\textbf{DeepGlobe}} is a dataset of satellite images for land cover classification, consisting of 6 classes. The image size is 512×512, downscaled to 224×224 for input. All three datasets are split into training, validation, and testing sets with a ratio of 80\% : 10\% : 10\%. Considering the uniform distribution of orbits and the even arrangement of satellites, these datasets are assumed to be independent and identically distributed.

\begin{figure}[t]
    \centering
    \includegraphics[width=\linewidth]{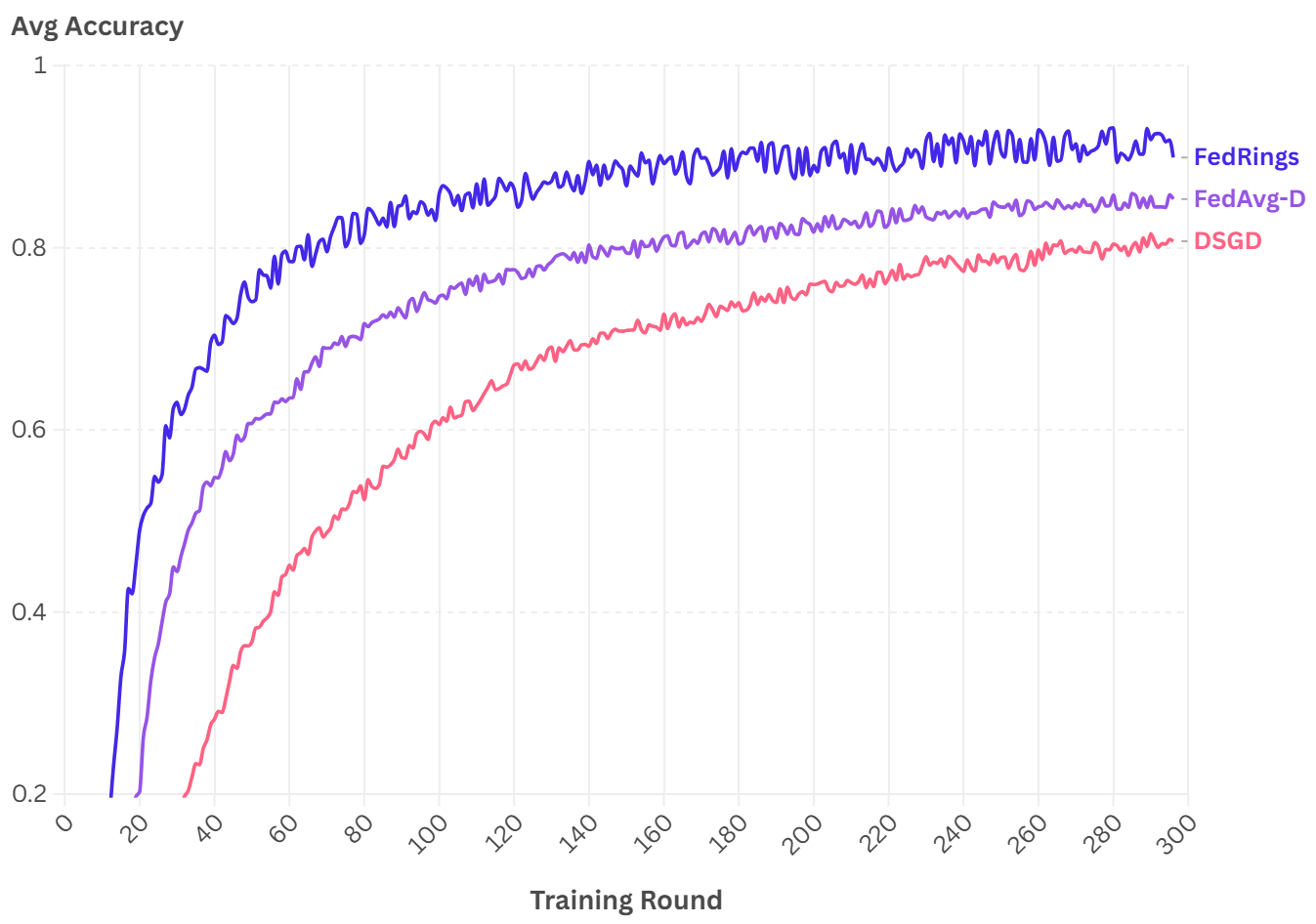}\\
    \caption{\small Test accuracy for EuroSAT.}
    \label{fig:EuroSAT}
\end{figure}

\begin{figure}[t]
    \centering
    \includegraphics[width=\linewidth]{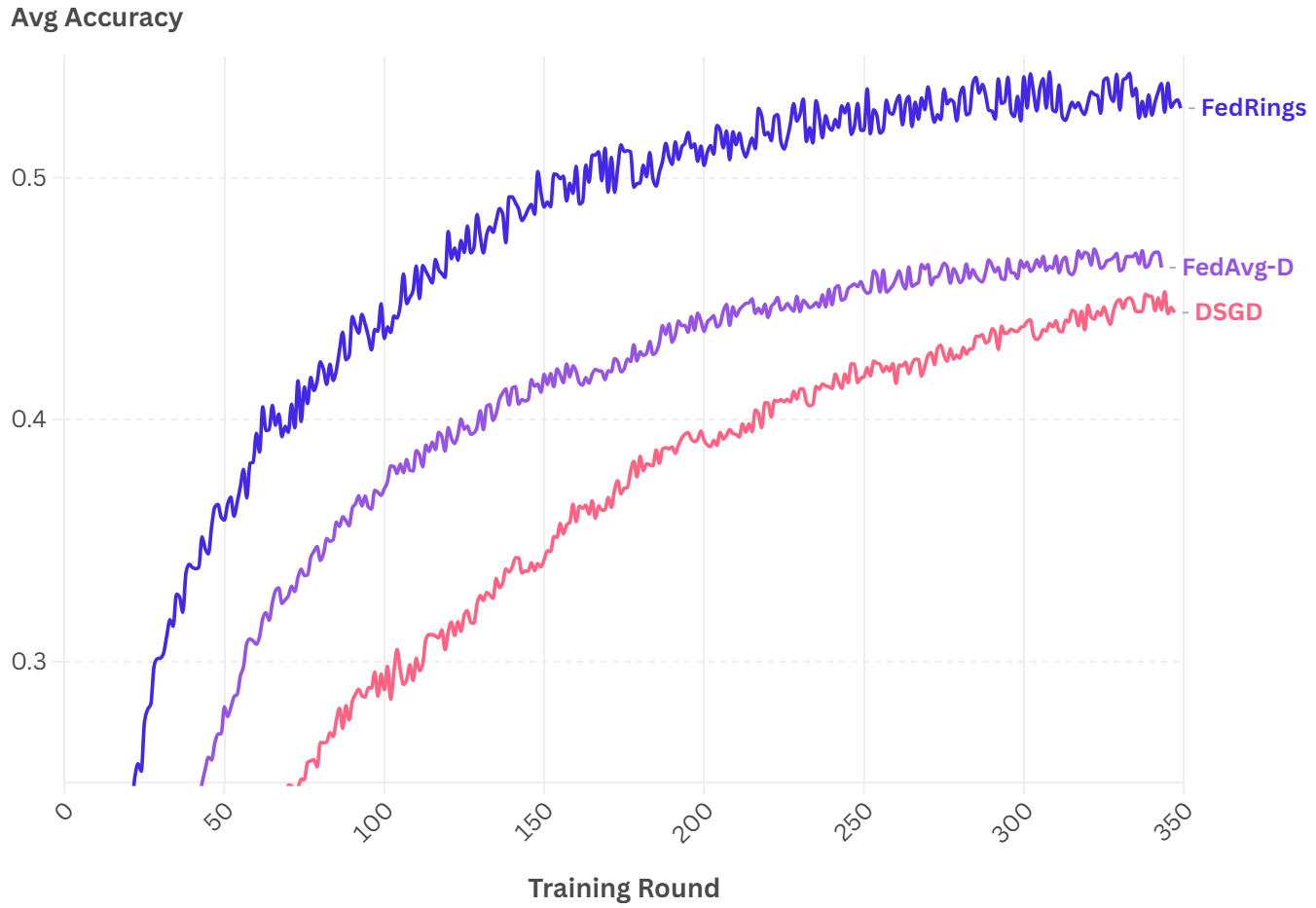}\\
    \caption{\small Test accuracy for So2Sat.}
    \label{fig:So2Sat}
\end{figure}

\begin{figure}[t]
    \centering
    \includegraphics[width=\linewidth]{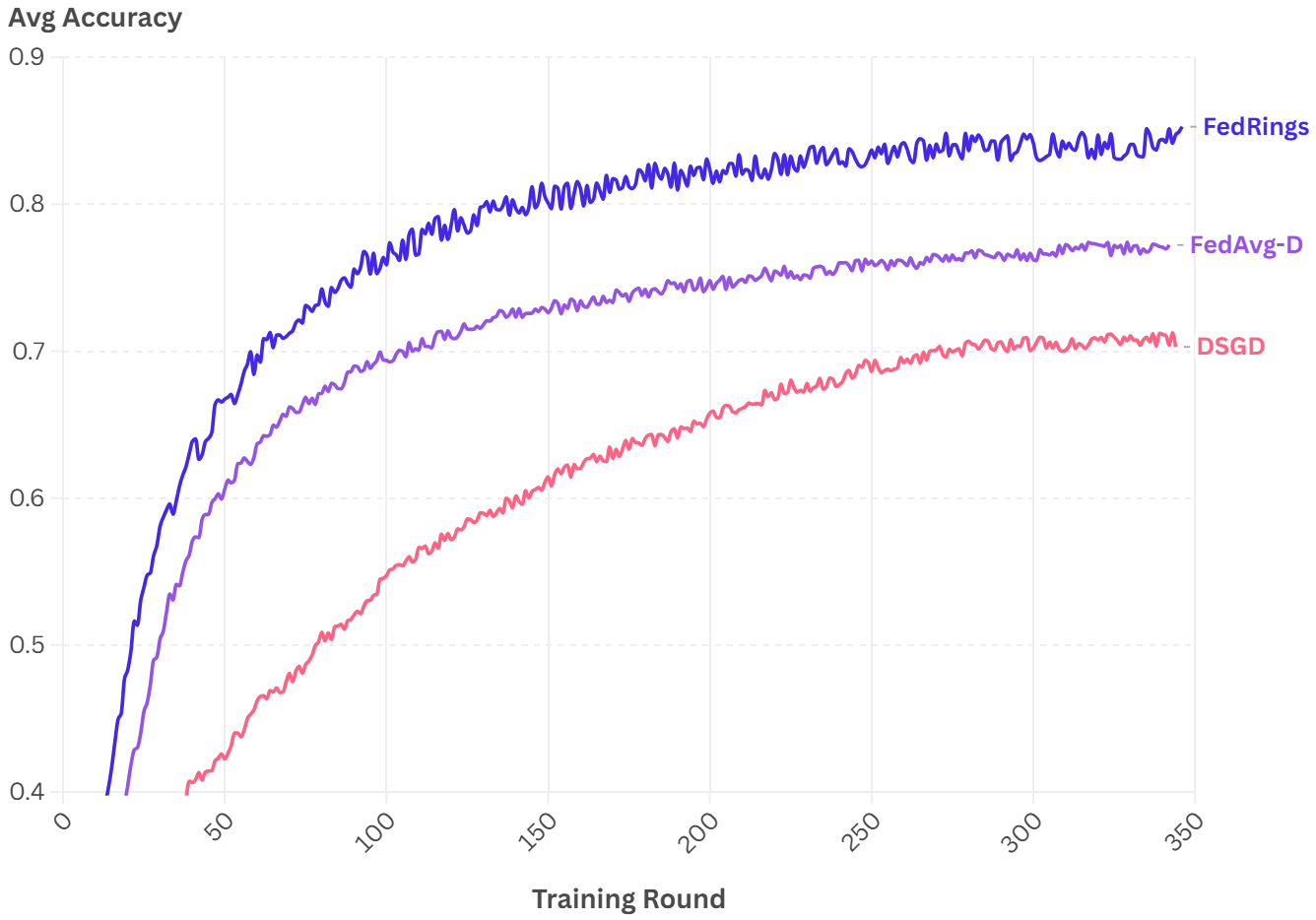}\\
    \caption{\small Test accuracy for DeepGlobe.}
    \label{fig:DeepGlobe}
\end{figure}

\subsection{Baseline}
\label{sec:baseline}
Although research on federated learning in satellite environments has gained increasing attention in recent years, as of now, there are no publicly available benchmark algorithms. Moreover, the 12 algorithms mentioned in the literature review of this paper are also not open-sourced. Therefore, the baselines chosen for this study are the relatively simple and basic FedAvg-D (a decentralized version
of FedAvg) and DSGD (a decentralized version of SGD). During the experiments, they were subjected to the same simulated signal interruptions and packet loss conditions as FedRings.

\subsection{Convergence Performance}
In convergence experiments, the performance of three different algorithms (FedRings, FedAvg-D, DSGD) on three different datasets was tested. As shown in Figures~\ref{fig:EuroSAT}, \ref{fig:So2Sat}, and ~\ref{fig:DeepGlobe}, the FedRings framework demonstrates significant advantages in convergence performance across the three datasets. On the EuroSAT dataset, FedRings achieves rapid convergence through the sparse incremental aggregation mechanism, reaching over 85\% accuracy within 50 rounds, far surpassing baseline algorithms. In the complex non-IID scenario of the So2Sat dataset, FedRings maintains robustness, with its routing optimization strategy effectively mitigating the impact of link disruptions, resulting in an accuracy improvement of over 5\%. On the DeepGlobe dataset, despite the increased training complexity due to high-resolution data, FedRings' sparsification strategy exhibits superior communication efficiency and convergence performance. These results demonstrate that FedRings excels in model training capability and adaptability under dynamic network and resource-constrained environments.

\begin{figure}[t]
    \centering
    \includegraphics[width=\linewidth]{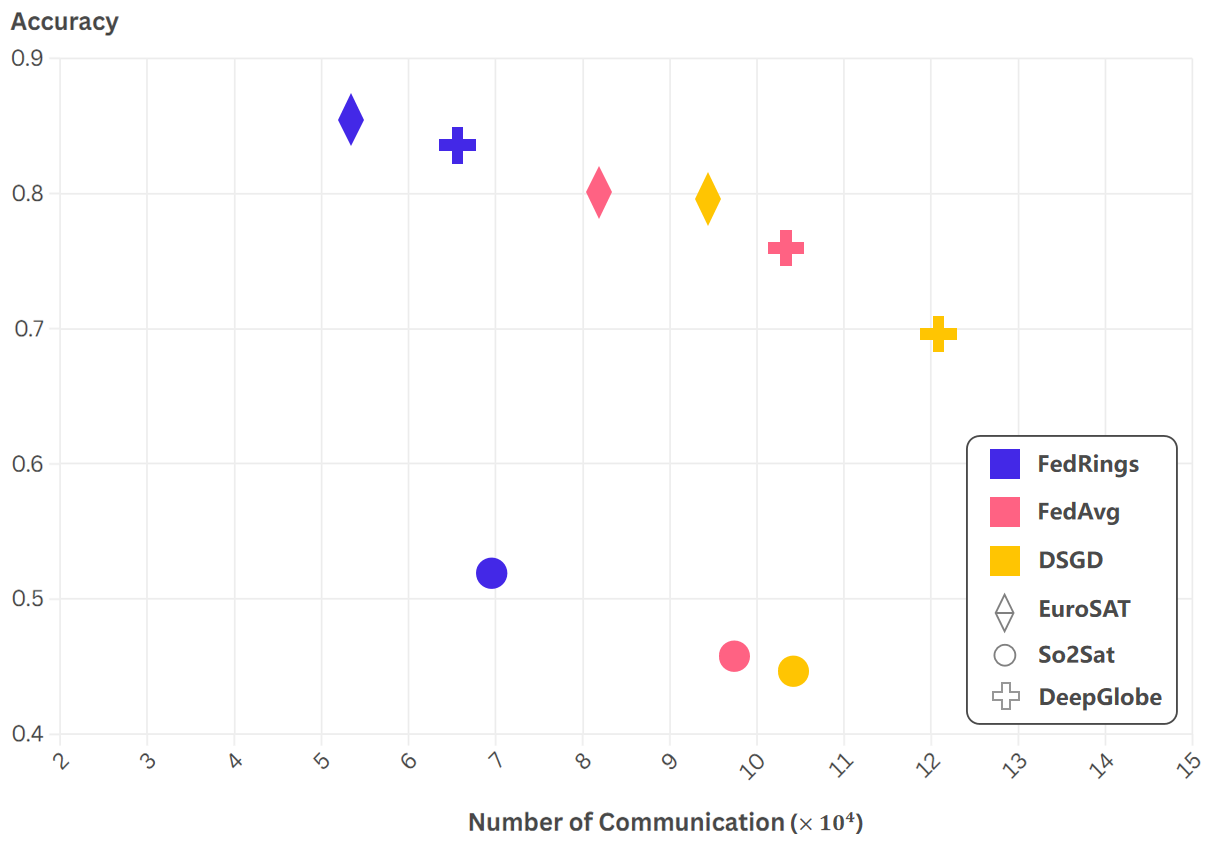}
    \caption{\small Communication overhead across 3 datasets using 3 algos.}
    \label{fig:communication}
\end{figure}

\begin{figure*}[t]
    \centering
    \begin{minipage}[b]{0.49\textwidth}
        \centering
        \includegraphics[width=\linewidth]{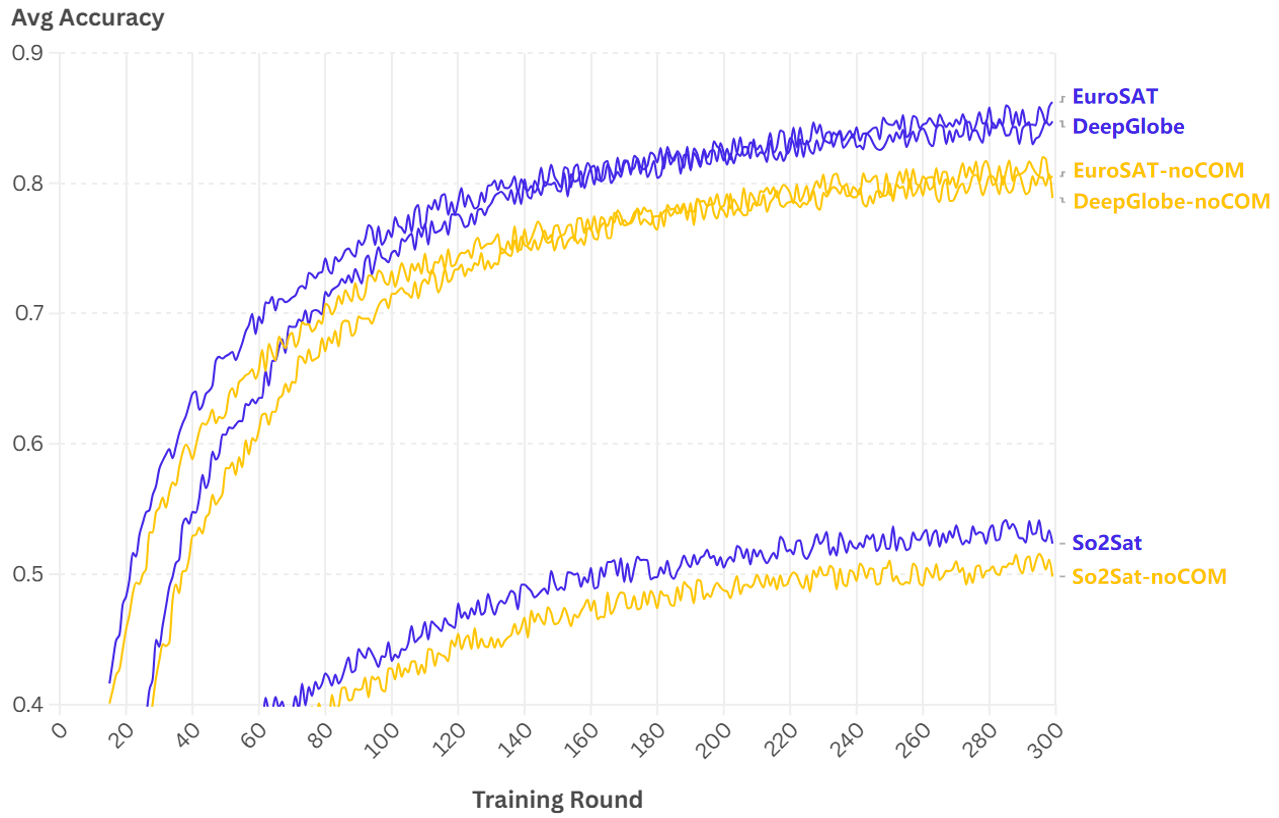}
        \caption{\small The influence of COM under 3 different datasets.}
        \label{fig:COM}
    \end{minipage}\hfill
    \begin{minipage}[b]{0.49\textwidth}
        \centering
        \includegraphics[width=\linewidth]{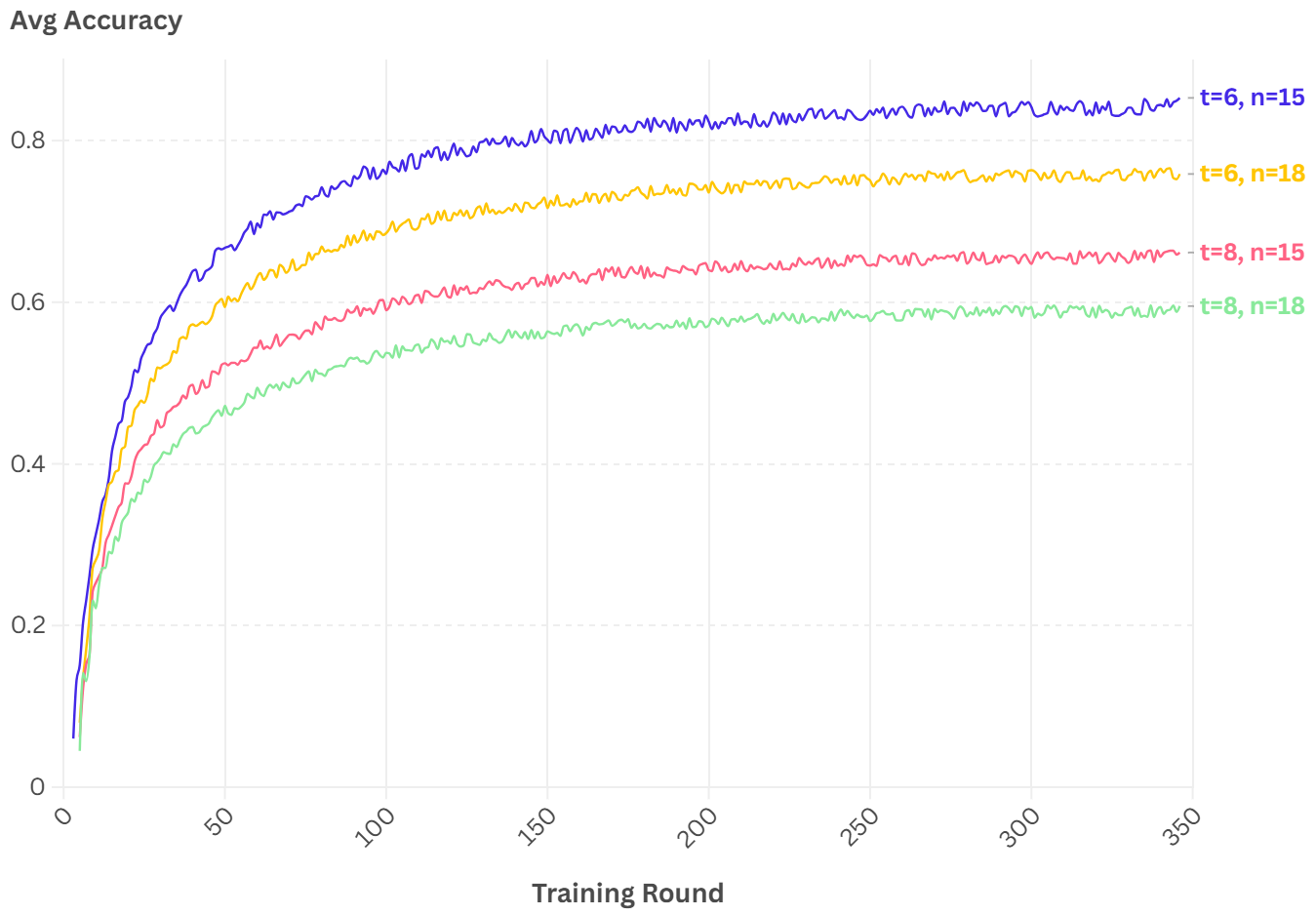}
        \caption{\small The Impact of Scalability on FedRings.}
        \label{fig:Scalability}
    \end{minipage}
\end{figure*}

\subsection{Communication Efficiency}
In the previously mentioned experiments, we recorded the total communication overhead along with the accuracy after completing all training rounds. In the evaluation of communication efficiency, as shown in Figure~\ref{fig:communication}, FedRings demonstrates significantly lower communication overhead across all three datasets (EuroSAT, So2Sat, and DeepGlobe) compared to baseline methods FedAvg-D and DSGD. Notably, in more complex datasets (So2Sat and DeepGlobe), FedRings' advantage becomes even more pronounced. The reduction in communication overhead is attributed to the ASIA mechanism. By utilizing Top-Q sparsification, FedRings transmits only the most critical parameters, drastically reducing redundant data. Furthermore, its incremental aggregation strategy ensures that each node transmits only aggregated updates rather than the entire parameter matrix. The time-correlated sparsification enhances this efficiency by maintaining stable key parameters across multiple rounds, minimizing unnecessary fluctuations. These results highlight FedRings' suitability for communication-constrained environments, such as LEO satellite constellations, where short communication windows and limited bandwidth are major challenges. The framework's ability to balance communication efficiency and model convergence makes it a robust choice for decentralized federated learning in dynamic satellite networks.

\subsection{COM Influence}
To understand the impact of spatio-temporal dynamic routing on FedRings, we attempted to demonstrate the differences in convergence performance by removing the COM. Figure~\ref{fig:COM} shows that removing COM leads to slower convergence across all datasets, particularly in the early training stages. The final accuracy also decreases noticeably, with more severe performance degradation on the more complex So2Sat and DeepGlobe datasets. The results demonstrate that COM plays a critical role in improving FedRings' convergence speed and accuracy. By predicting communication windows and optimizing routing paths, COM reduces delays and transmission failures, enhancing the efficiency and stability of model training. Without COM, FedRings' performance declines significantly, especially under unstable communication links or higher data complexity, highlighting its importance in dynamic satellite networks.

\subsection{Scalability}
To evaluate the scalability of FedRings, we explored the impact of increased scale on convergence efficiency by varying the number of satellites within each orbital plane and the total number of orbital planes in the constellation. Figure~\ref{fig:Scalability} illustrates that as the network scale increases, FedRings maintains stable convergence and accuracy compared to the baseline methods. However, the convergence rate slightly slows down as the number of satellites grows, particularly when both the
satellite count and orbital planes are increased simultaneously. Although larger-scale constellations introduce minor convergence delays, FedRings' robustness and efficiency remain superior, making it well-suited for dynamic and expansive satellite networks. The results confirm that FedRings is highly scalable and can adapt to larger satellite constellations while preserving model performance. Its design, including ASIA and spatio-temporal routing, ensures efficient parameter updates and communication even as the network size increases.




\begin{table*}[!t]
\centering
\caption{Comparison of Federated Learning Approaches in LEO Satellite Networks}
\label{tab:fl_leo_comparison}

\small
\setlength{\tabcolsep}{3pt}

\begin{tabular}{p{2.6cm} p{0.8cm} p{2.2cm} p{3.0cm} p{3.2cm} p{4.0cm}}
\hline
\textbf{Work} & \textbf{Year} & \textbf{Arch.} & \textbf{Key Idea} & \textbf{Strengths} & \textbf{Limitations / Gaps} \\
\hline

AsyncFLEO & 2022 & Hybrid & Async FL via HAPs & Handles intermittent links & Depends on HAP infrastructure \\

FedHAP & 2022 & Hybrid & FL via HAPs & Faster convergence & Infrastructure dependency \\

Ground-assisted FL & 2022 & Centralized & Ground aggregation & Simple coordination & High latency, poor scalability \\

FedSpace & 2022 & Sat-Ground & Joint FL & Practical deployment & Heavy ground reliance \\

DSFL & 2022 & Decentralized & Energy-aware FL & No central server & Simplified communication \\

FedGSM & 2023 & Decentralized & Staleness mitigation & Better convergence & No topology awareness \\

FedFusion & 2023 & Hybrid & Multi-modal FL & Handles heterogeneity & No comm optimization \\

FedSN & 2023 & General & Generic FL & Flexible design & Limited topology modeling \\

FedLEO & 2023 & Decentralized & Offloading & Resource efficiency & Coordination overhead \\

FedUR & 2023 & Central & Adaptive optimizer & Better convergence & Not sat-specific \\

Xu et al. & 2024 & Sat-Ground & Density-aware FL & Adapts to connectivity & Ground dependency \\

DFedSat & 2024 & Decentralized & ISL-based FL & Fully decentralized & No structured topology \\

FEDMEGA & 2024 & Sat-Ground & Contact scheduling & Comm-efficient & Limited ISL focus \\

OSC-FSKD & 2024 & Decentralized & Clustering FL & Handles non-IID & Ignores topology \\

Dataset Distill. & 2025 & Hybrid & Synthetic data & Reduces comm & Not topology-aware \\

ALANINE & 2025 & Decentralized & Personalized FL & Handles heterogeneity & No routing design \\

Sparse Incremental Aggregation & 2025 & Central & Incremental aggregation & Bandwidth efficiency & Rely on centralized aggregation \\

HiSatFL & 2025 & Hierarchical & Multi-layer FL & Scalable, privacy-aware & High complexity \\

RAFL & 2025 & Hierarchical & Reputation FL & Robust to attacks & Overhead, assumptions \\

Semi-sup. FL & 2025 & Hierarchical & Clustering + comp. & Efficient learning & Limited topology awareness \\

Blockchain FL & 2025 & Decentralized & Trust via blockchain & Secure & High overhead \\
sat-QFL & 2025 & Hierarchical & Quantum FL & Strong security & Practicality issues \\

\textbf{FedRings} & \textbf{2026} & \textbf{Decentralized} & \textbf{Ring-based FL} & \textbf{Scalable, no ground} & \textbf{Security} \\

\hline
\end{tabular}
\end{table*}

\subsection{Comparison with State of the Art}
Table~\ref{tab:fl_leo_comparison} summarizes federated learning approaches in LEO satellite networks. Most prior works either rely on centralized or hybrid architectures that focus on ground infrastructure (e.g., AsyncFLEO, FedHAP, FedSpace, FEDMEGA), or adopt decentralized approaches but lack topology awareness or oversimplify communication (e.g., DFedSat, FedGSM, DSFL). Hierarchical approaches such as HiSatFL, RAFL, and semi-supervised clustering-based FL, improve scalability but introduce significant complexity and overhead.

Early satellite FL works (AsyncFLEO, FedHAP, FedSpace) offload aggregation to HAPs or ground stations, achieving faster convergence at the cost of infrastructure dependency and high latency. DSFL emerged as an early decentralized alternative using energy-aware gossip communication, but with a simplified, topology-agnostic model. Later, FedGSM, FedLEO, FedSN, FedFusion, and FedUR pushed toward decentralization and resource efficiency, tackling gradient staleness, computation offloading, and optimizer adaptivity, yet none model the physical ISL topology explicitly. DFedSat, FEDMEGA and OSC-FSKD mark a shift toward practical ISL-based and density-aware designs, with DFedSat achieving full decentralization over ISLs and FEDMEGA optimizing contact scheduling, but structured topology modeling remains absent. Later works diversify further, with HiSatFL and RAFL introducing hierarchical architectures with privacy and reputation mechanisms at high coordination cost. Sparse Incremental Aggregation improves bandwidth efficiency but requires a central aggregator, while ALANINE targets personalization without routing support, and Blockchain FL and sat-QFL pursue security through consensus and quantum cryptography, respectively, each introducing substantial overhead or practical feasibility concerns, despite the security benefits.

In contrast, FedRings adopts a fully decentralized topology aware approach, that eliminates dependency on ground, while providing realistic spatio-temporal routing by means of the COM and ASIA. Compared to other decentralized works, FedRings, uniquely combines topology-aware routing, sparse incremental aggregation, and a historical compensation mechanism, while reducing overhead, accelerating model convergence and improving scalability. The primary remaining limitation is the absence of built-in security mechanisms, which we identify as future work.



\section{Conclusion and Future Work}
\label{sec:conclusion}
This paper presented FedRings, a FL framework designed specifically to address the unique challenges faced in LEO satellite constellations. The framework provides innovative solutions to issues such as intermittent connectivity, limited communication bandwidth, and highly dynamic network topologies. By leveraging topology-aware spatio-temporal routing strategies, ASIA and a historical compensation mechanism, FedRings enhances communication efficiency, model synchronization, and robustness. FedRings' ability to significantly reduce communication overhead, accelerate model convergence, and scale efficiently in complex LEO environments, makes it a practical and impactful solution.

Future work includes the extension of FedRings to MEO and GEO constellations, the exploration of more adaptive parameter aggregation techniques and hybrid synchronization mechanisms that balance efficiency and stability, as well as the incorporation of advanced spatio-temporal routing strategies, leveraging machine learning models to predict satellite orbits and optimize communication windows more effectively. 
Furthermore, as mega-constellations 
continue to grow, FedRings could be adapted to handle larger networks by implementing hierarchical aggregation models and clustering strategies. These advancements would not only enhance FedRings' scalability but also broaden its applicability to a wide range of space-based distributed learning scenarios. Finally, a security-centered approach to FedRings constitutes another relevant line of work for the future.

\newpage






\bibliographystyle{plain}
\bibliography{fedrings}

\end{document}